\documentclass[12pt]{article}
\usepackage{amssymb}
\usepackage{hyperref}
\usepackage[dvips]{epsfig}
\usepackage[utf8]{inputenc}

\newcommand{\be}{\begin{equation}}
\newcommand{\ee}{\end{equation}}

\newcommand{\bn}{\begin{eqnarray}}
\newcommand{\en}{\end{eqnarray}}

\newcommand{\p}{\partial}

\def\bea{\begin{eqnarray}}
\def\eea{\end{eqnarray}}

\newcommand{\beq}{\begin{eqnarray}}
\newcommand{\eeq}{\end{eqnarray}}

\usepackage[utf8]{inputenc}
\usepackage[dvips]{geometry}

\usepackage{indentfirst}
\usepackage{amsmath}
\usepackage{graphicx}
\usepackage{amssymb}
\def\({\left(}
\def\){\right)}

\def \= {\,\dot{=}\,}
\begin{document}

\title{\textbf{Duality and Self-duality of the Spin-1 Model in the Covariant Operator Formalism}}
\author{ 
G. B. de Gracia\footnote{gb9950@gmail.com},\ \quad B. M. Pimentel \footnote{b.m.pimentel@gmail.com},\ \quad L. Rabanal \footnote{luis.rabanal@unesp.br} \\ 
\textit{{Instituto de F\'{i}sica Te\'{o}rica (IFT), Universidade Estadual Paulista (UNESP)} }\\
\textit{{Rua Dr. Bento Teobaldo Ferraz, 271,
Bloco II, Barra Funda} }\\
\textit{{CEP 01140-070-S\~{a}o Paulo, SP, Brazil} } \\}
\date{}
\maketitle

\begin{abstract}
We perform the covariant operator quantization of the spin-$1$ model in $2+1$ spacetime dimensions to rigorously establish its dualities. For this purpose, the Kugo-Ojima-Nakanishi formalism, based on an indefinite metric Hilbert space in the Heisenberg picture, is used. We show that it is possible to extract a massive physical excitation constructed from a linear combination of the vector field $A_{\mu}$ and the $B$-field. In turn, we also show that this excitation generates the Maxwell-Chern-Simons theory. This is achieved by exploring the two-point function of the vector field.
\end{abstract}

\section{Introduction}
Quantum field theories describing massless vector fields in low dimensionality exhibit, besides the traditional ultraviolet (UV) divergences, severe infrared (IR) singularities. In the early eighties, a topological mass term was introduced to provide an infrared cut-off for these theories in $2+1$ spacetime dimensions without neither violating gauge symmetry nor adding new degrees of freedom \cite{deser1}. The resulting theory is the so-called Maxwell-Chern-Simons (MCS) theory which describes a topological massive excitation. A nice review of these ideas is given in reference \cite{dunne}.

After that, a self-dual spin $1$ model was introduced by Townsend and collaborators \cite{townsend}. Even though this new theory describes a massive vector field without imposing gauge symmetry, it was realized that the model is related to the above topological massive model through a dual correspondence. 

The discovery of duality relations among theories seemingly distinct from each other had had important applications in physics ranging from well-established weak dualities in condensed matter \cite{kramers,peskin} to conjecture dualities, e.g., a duality web \cite{webduality1,webduality2} or the AdS/CFT correspondence \cite{maldacena}. Hence, dualities in general deserve to be investigated further.

The above-mentioned equivalence between the self-dual model and the MCS theory was shown to hold even in the presence of external sources \cite{deser} and it has also been studied in the Hamilton-Jacobi formalism \cite{pimentel}. Yet, in this work we provide an alternative description of the duality using the covariant operator or Kugo-Ojima-Nakanishi (KON) formalism \cite{kugoojima,Nakanishi}. As previously explained, the self-dual model is a first-order theory that describes massive spin $1$ excitations without imposing gauge symmetry. Despite this, the KON formalism is suitable to perform a covariant quantization free of patologies, e.g., the smooth zero mass limit, in analogy with the Proca theory case \cite{Nakanishi}. 

The paper is organized as follows. In section $2$, we show that it is possible to find a physical massive excitation for the self-dual model which is a linear combination of the vector and the auxiliary $B$-field. Section $3$ we infer that the massive mode two-point function of the self-dual theory is the same as that of the positive norm excitation of the MCS model \cite{sete,gabriel}. This will be our derivation of the duality. It is interesting to mention that although the MCS model has gauge symmetry, while the self-dual model has not, it act just on non-physical modes allowing a map between their observable sectors. Our conclusions are given in section $4$. The metric signature $+ - -$ is used throughout.

\section{The self-dual spin 1 model in $2+1$ dimensions}
We start from the self-dual Lagrangian coupled to the auxiliary $B$-field
\begin{equation}
    {\cal{L}_{\text{SD}}} = \frac{m^2}{2}A_\mu A^\mu-\frac{m}{2}\epsilon_{\alpha \beta \gamma}A^\alpha \p^\beta A^\gamma +B\p_\mu A^\mu+\frac{\alpha B^2}{2}.
\end{equation}
The canonical momenta of the fields are respectively
\begin{align}
\pi_i&=\frac{\p {\cal{L}} }{\p(\p_0 A_i)}=-\frac{m}{2}\epsilon^{ij}A_j \\ \nonumber \pi_0&=\frac{\p {\cal{L}} }{\p(\p_0 A_0)}=B \\ \nonumber
\pi_B&=\frac{\p {\cal{L}} }{\p(\p_0 B)}=0.
\end{align}

The non-vanishing Dirac brackets are \cite{oito}
\begin{equation}
    \{ A_{\mu}(x), \pi^\nu(y) \}= \frac{1}{2} \delta_\mu^\nu \delta^2(x-y).
\label{overallfactor}
\end{equation}
Since the factor of $1/2$ cannot be trivially obtained, we shall present its derivation in the Appendix. The application of the correspondence principle implies that the canonical equal-time commutators have the following form 
\begin{align}
\big[A_i(x),A_j(y) \big]_0 &=-i\frac{\epsilon_{ij}}{m}\delta^2(x-y) \label{AiAj}\\
\big[A_0(x),B(y) \big]_0 &=i\delta^2(x-y), \label{A0B}
\end{align}
with the subscript $0$ meaning equal time, that is, $x_0 = y_0$.

\subsection*{Equations of motion}
The equation of motion for the vector field operator reads
\begin{equation}
    m^2A_\mu-m\epsilon_{\mu \alpha \beta}\p^\alpha A^\beta=\p_\mu B,
\label{vectoreom}
\end{equation}
whereas for the $B$-field we have
\begin{equation}
    \p_\mu A^\mu=-\alpha B.
\label{gaugecondition}
\end{equation}
If we take the divergence of equation \eqref{vectoreom} and use the gauge condition \eqref{gaugecondition}, we find that the $B$-field satisfies
\begin{equation}
    \left(\Box+\alpha m^2\right) B=0. 
\label{EqForB}
\end{equation}
Hence, as usual in the case of an Abelian theory, the subsidiary condition necessary to identify the physical space $\mathfrak{F}_{\text{phys}}$ is given by
\begin{equation}
B^+(x) | \text{phys} \rangle = 0, \quad \forall | \text{phys} \rangle \in \mathfrak{F}_{\text{phys}},
\label{subsidiarycondition}
\end{equation}
where $B^+(x)$ denotes the positive frequency part of the solution of the elliptic differential operator equation \eqref{EqForB}.

\subsection*{Initial conditions}
The gauge condition (\ref{gaugecondition}) can be used to derive the following initial condition
\begin{equation}
    [A_0(x),\dot A_0(y)]_0=-i\alpha \delta^2(x-y).
\end{equation}
The $\mu=0$ component of the equation (\ref{vectoreom}) together with the equal-time commutator (\ref{A0B}) gives
\begin{equation}
    \big[ B(x),\dot B(y) \big]_0=-im^2\delta^2(x-y).
\label{Binitial}
\end{equation}
Using again the $\mu=0$ component of equation (\ref{vectoreom}) as well as the canonical equal-time commutators, we can also obtain
\begin{equation}
    \big[ A_i(x),\dot B(y) \big]_0=-i\p_i\delta^2(x-y).
\label{AiB0}
\end{equation}

\subsection*{General solution of commutators}
The harmonic character of the $B$-field together with its integral representation \cite{Nakanishi}, namely,
\begin{equation}
B(y)=\int d^2z\big[\p_0^z\Delta(y-z;\alpha m^2)B(z)-\Delta(y-z;\alpha m^2)\p_0^zB(z)  \big],
\label{integralrepresentationB}
\end{equation}
and the equal-time commutation relations (\ref{Binitial}) gives
\begin{equation}
    \big[ B(x),B(y) \big]=-im^2\Delta(x-y;\alpha m^2).
\label{BBGeneral}
\end{equation}
Moreover, using the integral representation of the $B$-field (\ref{integralrepresentationB}) and the equal-time commutator (\ref{AiB0}) we have
\begin{equation}
    \big[A_\mu(x), B(y) \big]=-i\p_\mu \Delta(x-y;\alpha m^2).
\label{ABGeneral}
\end{equation}
where $\Delta(x-y;s)$ is the Green function of the harmonic equation (\ref{EqForB}) and its precise definition is given in equation (\ref{CauchyGreen}) below.

\subsection*{Physical massive excitation}
If one acts on the vector field equation (\ref{vectoreom}) with the differential operator $\epsilon^{\beta \sigma \mu}\p_\sigma$, consider the equation of motion for the $B$-field and the transversality of $A_\mu$, then, a purely massive excitation can be found explicitly to be of the form
\begin{equation}
    \big(  \Box+m^2\big){\cal{U}}^\mu(x)=0,\qquad \p_\mu\ {\cal{U}}^\mu(x)=0,
\end{equation}
with
\begin{equation}
{\cal{U}}^\mu(x)=A^\mu(x)-\frac{\p^\mu B(x)}{m^2}.
\end{equation}
Furthermore, we can use the general solution of the commutators (\ref{BBGeneral}) and (\ref{ABGeneral}) to compute
\begin{equation}
    \big[{\cal{U}}^\mu(x),B(y) \big]=0.
\end{equation}

This equation ensures that the massive excitation $\mathcal{U}^{\mu}(x)$ is indeed physical in the sense of the subsidiary condition (\ref{subsidiarycondition}). In the next section, we shall compute the commutator of this physical massive excitation, verify the self-duality and provide the map between the vector field two-point function with the purely massive MCS excitation.

\section{Two-point function, self-duality and the dual map}
Although an exact or non-perturbative answer for the commutator between vector fields at two different spacetime points in the presence of interactions is almost impossible, the spectral representation method helps us to extract valuable information. In particular, it guides the construction of the asymptotic fields of the theory which represent the in/out Fock spaces $\mathfrak{F}$.
As in the previous section, equal-time commutation relations, quantum equations of motion and symmetries is all what we need.

Using the equations of motion and the commutator for the $B$-field, we find a relation for the gauge field commutator
\begin{equation}
\left( m\eta^{\mu\alpha} - \epsilon^{\mu\sigma\alpha}\p_\sigma^x \right) \left( m\eta^{\nu\beta} - \epsilon^{\nu\gamma \beta}\p_\gamma^y \right) \big[A_\alpha(x),A_\beta(y)\big] = -i\p^\mu_x \p^\nu_y\Delta(x-y;\alpha m^2),
\end{equation}
for which the general solution is given by
\begin{equation} \big[A_{\mu}(x),A_\nu(y) \big] = ib\left(\eta_{\mu \nu}+\frac{\p_\mu \p_\nu}{m^2} - \frac{\epsilon_{\mu \nu \beta}\p^\beta}{m} \right)\Delta(x-y;m^2)
-\frac{i}{m^2}\p_\mu \p_\nu \Delta(x-y,\alpha m^2),
\end{equation}
wherein the operator $\Delta(x-y;s)$ is the massive Green function of D'Alembert equation which is defined by the following Cauchy data
\begin{equation} 
\Box \Delta(x-y; s) = -s\Delta(x-y; s), \quad \Delta(x-y; s)|_0 = 0, \quad \p_0^x\Delta(x-y; s)|_0 = -\delta^2(x-y). 
\label{CauchyGreen}
\end{equation}

The undetermined constant is found by using equal-time commutators. In order to have a vector field commutator consistent with $\big[A_{i}(x),A_j(y) \big]_0$ given in equation \eqref{AiAj}, we must have $b=-1$. We can see that $\big[A_{0}(x),\p_0A_0(y) \big]_0=-i\alpha\delta^2(x-y)$ is also in agreement with our solution. Consequently, the vector field commutator at unequal times is given by
\begin{align} 
\big[A_{\mu}(x),A_\nu(y) \big] &= -i\left(\eta_{\mu \nu} +\frac{\p_\mu \p_\nu}{m^2} - \frac{1}{m}\epsilon_{\mu \nu \beta}\p^\beta  \right)\Delta(x-y;m^2) \nonumber \\
&\quad-\frac{i}{m^2}\p_\mu \p_\nu \Delta(x-y;\alpha m^2).
\label{AA}
\end{align}

We can also conclude that the purely massive physical combination has the following two-point structure
\begin{equation}
    \big[{\cal{U}}_\mu(x),{\cal{U}}_\nu(y) \big] = -i\left(\eta_{\mu \nu} +\frac{\p_\mu \p_\nu}{m^2} - \frac{1}{m}\epsilon_{\mu \nu \beta}\p^\beta  \right)\Delta(x-y;m^2),
    \label{UU}
\end{equation}
since the following relation holds
\begin{equation}
    \big[A_{\mu}(x),A_\nu(y) \big] = \big[{\cal{U}}_\mu(x),{\cal{U}}_\nu(y) \big] - \frac{i}{m^2}\p_\mu \p_\nu \Delta(x-y;\alpha m^2).
\end{equation}
We realized that the commutator for the ${\cal{U}}_\mu(x)$ field has the same exact structure as that of the physical excitation of the MCS theory \cite{sete,gabriel} that we denote it by ${\cal{V}}_\mu(x)$. If, on the one hand, one applies the operator $\epsilon^{\alpha  \sigma \mu}\p_\sigma/m$ to (\ref{UU}) twice we have
\begin{equation}
    \left[\frac{1}{m}\epsilon^{\alpha \sigma \mu}\p_\sigma{\cal{U}}_\mu(x),\frac{1}{m}\epsilon^{\beta\gamma \nu}\p_\gamma{\cal{U}}_\nu(y) \right] = -i\left(\eta^{\alpha\beta} +\frac{\p^\alpha \p^\beta}{m^2} - \frac{1}{m}\epsilon^{\alpha\beta \xi}\p_\xi  \right)\Delta(x-y;m^2),
\end{equation}
and the same commutator is recovered, thus, proving its self-dual character. On the other hand, if one applies this same operator to (\ref{AA}) twice we have
\begin{equation}
    \left[\frac{1}{m}\epsilon^{\alpha \sigma \mu}\p_\sigma A_\mu(x),\frac{1}{m}\epsilon^{\beta\gamma \nu}\p_\gamma A_\nu(y) \right] = - i\left(\eta^{\alpha\beta} +\frac{\p^\alpha \p^\beta}{m^2} - \frac{1}{m}\epsilon^{\alpha\beta \xi}\p_\xi  \right)\Delta(x-y;m^2).
\end{equation}
Therefore, the resulting commutator becomes equal to the physical MCS excitation ${\cal{V}}_\mu(x)$, thus, establishing the duality. 

Summing up, we have derived the dual map and the self-duality relations
\begin{equation}
    \frac{1}{m}\epsilon_{\mu  \sigma \gamma} \p^\sigma A^\gamma \longleftrightarrow {\cal{V}}_\mu, \qquad \frac{1}{m}\epsilon_{\mu  \sigma \gamma} \p^\sigma {\cal{U}}^\gamma  \longleftrightarrow {\cal{U}}_\mu.
\end{equation}
This shows that the dualities found between those models by means of path integral methods has indeed a canonical quantization counterpart as it should be.

\section{Conclusion}
We have shown that the KON formalism gives an alternative approach to find the duality matching between a theory with gauge symmetry (MCS theory) and another without it (self-dual model). Although we have pointed out that this duality has already been studied by using the Hamilton-Jacobi formalism and other methods, we believe that the importance of the KON formalism lies in the fact that the Heisenberg operator quantization is the most fundamental one.

The method allowed us to separate physical sectors from the gauge ones. In this sense, we have been able to show the existence of a purely massive physical excitation $\mathcal{U}_{\mu}(x)$ formed by a linear combination of the vector field and the $B$-field. Its commutator had the same exact structure as that of the physical MCS excitation and we have provided a prove for its self-dual character. It became clear that the map taking the self-dual model to the MCS theory eliminates the $\alpha$-dependent sector of the former given by the last term in \eqref{AA}, thus, extracting only the physical part as it should be.  

This work leaves open the possibility of the study of more dualities within the KON formalism. It could even help to rigorously establish previously conjecture dualities \cite{webduality1,webduality2} if appropriate extensions of the formalism were performed. We leave these studies for the future.

\section*{Acknowledgments}
G. B. de Gracia and L. Rabanal thank CAPES for support, and B. M. Pimentel thanks CNPq for partial support.

\appendix

\section{Appendix}
In this appendix we shall derive explicitly the Dirac brackets to justify the overall factor in equation (\ref{overallfactor}). The KON formalism fixes the action in such a way that there are no first-class constraints ambiguities, thus, the Dirac brackets can be used from the outset. In order to see this in our case of interest, let us build a matrix with all the Poisson brackets of the Legendre transform constraints
\begin{equation} 
M^{IJ}(x,y) = \left\{ \Phi^I(x),\Phi^J(y) \right\} = \left(\begin{matrix}
\epsilon^{nk}m& 0& 0       \\
0 & 0 & 1       \\
0& -1 & 0        
\end{matrix}\right) \delta^2(x-y),  
\end{equation}
where 
\begin{equation}
    \Phi^I(x) = \left(\begin{matrix}
\pi_i(x) + \frac{m}{2}\epsilon_{ij}A^j(x)\\
\pi_B(x) \\
\pi_0(x)-B(x)
\end{matrix}\right).
\end{equation}
The inverse matrix is given by
\begin{equation}
    \tilde{M}^{IJ}(x,y) = \left(\begin{matrix}
\frac{\epsilon^{nk}}{m}& 0& 0  \\
0 & 0 & 1     \\
0& -1 & 0     
\end{matrix}\right) \delta^2(x-y).
\end{equation}
Therefore, the reduced brackets are
\begin{align} 
\{A_i(x),\pi_j(y)\}_D &= \{A_i(x),\pi_j(y)\} + \int d^3w d^3z \{A_i(x),\Phi^1_n(w)   \}\frac{\epsilon_{mn}}{m}\{\Phi^1_m(z),\pi_j(y)   \} \nonumber \\
&= \frac{\delta_{ij}}{2}\delta^2(x-y),
\end{align}
instead of simply $\delta_{ij}\delta^2(x-y)$. This result allows us to derive the correct factor for the equal-time commutator between the spatial components of the vector field.

\end{document}